\documentclass[prb,twocolumn,superscriptaddress]{revtex4}
\usepackage{amsmath}
\usepackage{amssymb}
\usepackage{graphicx}

\newcommand{\commentout}[1]{}

\newcommand{\eref}[1]{eqn.~(\ref{#1})}
\newcommand{\fref}[1]{fig.~\ref{#1}}

\newcommand{\sref}[1]{sec.~\ref{#1}}

\newcommand{\frefs}[2]{figs.~(\ref{#1},\ref{#2})}

\newcommand{\Fref}[1]{Fig.~\ref{#1}}

\newcommand{\etal}{\emph{et al.}}

\newcommand{\cre}[1]{c_{#1}^\dagger}

\newcommand{\dcre}[1]{d_{#1}^\dagger}
\newcommand{\dann}[1]{d_{#1}^{\phantom\dagger}}
\newcommand{\ddes}{\dann}

\newcommand{\D}{\mathrm{d}}

\newcommand{\bk}{\mathbf{k}}
\newcommand{\K}{\bk}

\newcommand{\I}{\mathrm{i}}

\renewcommand{\Im}{\mathop{\mathrm{Im}}}

\newcommand{\ie}{\emph{i.e.}}
\newcommand{\cf}{\emph{cf.}}
\newcommand{\eg}{\emph{e.g.}}

\newcommand{\via}{\emph{via}}

\newcommand{\mgcomment}[1]{}
\newcommand{\boxcomment}[1]{}

\newcommand{\pd}{\phantom\dagger}

\newcommand{\half}{\tfrac{1}{2}}

\begin{document}

\author{Martin R. Galpin}
\author{Frederic W. Jayatilaka}
\author{David E. Logan}
\affiliation{Chemistry Department, Oxford University, Physical \& Theoretical Chemistry, South Parks Road, Oxford  OX1 3QZ, UK}

\author{Frithjof B. Anders}
\affiliation{Technische Universit\"at Dortmund, Theoretische Physik II,
 44221 Dortmund, Germany}

\date{November 11, 2009}
\title{Interplay between Kondo physics and spin-orbit coupling \\ in carbon nanotube quantum dots}

\begin{abstract}
We investigate the influence of spin-orbit coupling on the Kondo effects in carbon nanotube quantum dots, using the
numerical renormalization group technique. A sufficiently large spin-orbit coupling is shown to destroy the $SU(4)$ Kondo effects at zero magnetic field, leaving only two $SU(2)$ Kondo effects in the one- and three-electron Coulomb blockade valleys. On applying a finite magnetic field, two additional, spin-orbit induced $SU(2)$ Kondo effects  arise in the three- and two-electron valleys. Using physically realistic model parameters, we calculate the differential conductance over a range of gate voltages, temperatures and fields. The results agree well with measurements from two different experimental devices in the literature, and explain a number of observations that are not described within the standard framework of the $SU(4)$ Anderson impurity model. 
\end{abstract}

\maketitle

\section{Introduction}
\label{sec:intro}
Quantum dots fabricated within carbon nanotubes (CNTs) have attracted considerable attention in recent years (for reviews see \eg \ [\onlinecite{Dekker1999,Sapmaz2006}]). Such devices, in which electrons are trapped within a small, strongly interacting region of the CNT by an applied electric field, show remarkable electronic transport properties\cite{Dekker1999,Sapmaz2006} and may have useful applications in future technology.\cite{Avouris2007} 

CNT-based devices are of particular interest due to their doubly-degenerate orbital structure which, combined with electron spin, generates a wealth of basic physics. One such phenomenon is the Kondo effect,\cite{hewsonbook} resulting from strong electron interactions within the dot. It is observed\cite{nygard,Buitelaar2002,jarillo,jarillo2,FinkelsteinI2007,FinkelsteinII2007} when the device is tuned so that the dot has a partially-filled shell of electrons: on lowering the temperature, the dot's 
spin/orbital degrees of freedom become strongly coupled\cite{hewsonbook} to those of the leads, leading to a complex many-body ground state with an enhanced electronic conductance.\cite{nglee,glazmanraikh} Understanding this effect is particularly important for CNT dots, because the involvement of both spin and orbital degrees of freedom generates an $SU(4)$ Kondo effect that persists to considerably higher temperatures -- up to a few Kelvin (\ie \ a few tenths of an $\text{meV}$) -- than the standard $SU(2)$ Kondo effect in semiconductor devices.\cite{goldhaber,cronenwett,vanderwiel} Using a range of many-body techniques, the theory of the $SU(4)$ Kondo effect is now well
established\cite{borda,Boese2002,zarand,LeHur2004,Galpin2005,Lopez2005,Choi2005,Galpin2006a,Galpin2006b,Mitchell2006,Rui2006,Sakano2006,Lim2006,LeHur2007,Busser2007,Anders2008,Mizuno2009} and \eg\ has been shown\cite{Anders2008} to be in good agreement with experiments\cite{FinkelsteinII2007} performed in the absence of a magnetic field.

Another consequence of the interplay between spin and orbital degrees of freedom in CNT devices is spin-orbit (SO) coupling. Its existence was beautifully demonstrated in experiments\cite{rmce} on a very strongly correlated
CNT dot, where it generates a splitting of sequential tunneling spectra at finite bias, and kinks in the magnetic field dependence of the Coulomb blockade `staircase' at zero bias. The strength of the SO coupling was measured\cite{rmce}, and for the device studied found to be of order $0.2$--$0.4 ~\text{meV}$, varying somewhat between different electron shells. 

Comparing the typical energy scales of the Kondo effect and the SO coupling, it is striking that \emph{both} may arise on the energy scale of a few tenths of an $\text{meV}$. Two related questions then arise: what effect does SO coupling have on the standard $SU(4)$ Kondo theory, and is the Kondo/SO competition seen experimentally? We seek to answer these questions in this paper.

Aspects of spin-orbit coupling in CNTs have recently been considered theoretically. The origins of the coupling itself have been determined from direct microscopic calculation,\cite{Ando,Heurtas} showing that while a number of distinct spin-orbit interactions arise in principle, the dominant contribution is the direct coupling between each electron's spin and orbital angular momentum. The effect of this coupling on the states of the isolated dot has been analysed in detail,\cite{Secchi2009,Wunsch2009} and the resulting sequential tunneling transport properties (arising when the Kondo scale is too small to be seen experimentally\cite{rmce}) have been calculated and compared to experiment.\cite{ourJCP} Aspects of the competition between spin-orbit coupling and the Kondo effect have also been studied\cite{TieFeng2008} via an equation of motion decoupling scheme in the $U\to\infty$ limit. 
Within this rather crude approximation,\cite{footnoteEoM} the splitting of the $SU(4)$ Kondo resonance was examined for finite SO coupling and magnetic field, and an orbital Kondo effect found at finite field in the two-electron Coulomb blockade valley.\cite{TieFeng2008}

In the present work we consider the two-fold orbitally degenerate $SU(4)$ Anderson impurity model (AIM) in a magnetic field, with SO coupling, and study it using the numerical renormalization group (NRG)~\cite{kww1,NRGrmp}
backed up by simple physical arguments. NRG is ideally suited to the problem, being known for similar quantum impurity models to provide numerically exact results on the low-energy/temperature scales relevant to experiment.
The model itself also has a strong track-record, a previous NRG study\cite{Anders2008} of the $SU(4)$ AIM in the absence of a magnetic field having shown that the all-important low-energy Kondo physics is well reproduced when the bare model parameters are fitted to high-energy conductance features such as the Coulomb blockade diamonds.

The paper is laid out as follows. The model is described in \sref{sec:model}, and the relevant theoretical background and NRG technique are discussed in \sref{sec:theory}. The behavior of the model in the atomic (lead-uncoupled dot) limit is outlined in \sref{sec:atomic},  from which simple arguments are then used to deduce the effect of introducing a finite spin-orbit interaction. The main body of the paper is \sref{sec:results}, where we present and discuss the results of NRG calculations. We begin by considering the zero-bias conductance, as a function of gate voltage, temperature and magnetic field. Here we make comparison to the experiments of Jarillo-Herrero \etal,\cite{jarillo2} showing that the orbital splitting identified empirically in experiment is readily explained by the inclusion of spin-orbit coupling in the model. We then turn to a discussion of the finite-bias conductance, comparing explicitly to the experiments of Makarowski \etal\cite{FinkelsteinI2007} and showing that the asymmetry observed in the Kondo peaks at finite bias is also well-described by the theory. 
The paper concludes with a brief summary, and a discussion of the applicability of the pure $SU(4)$ AIM to CNT quantum dots.

\section{Model}
\label{sec:model}
The basic model used to describe a CNT quantum dot is the $SU(4)$ Anderson impurity model\cite{Choi2005,Busser2007,TieFeng2008,Anders2008}, given in conventional notation by~\cite{hewsonbook}
\begin{equation}
\begin{split}
\label{eq:su4ham}
\hat H_{SU(4)} =& \sum_{\K,m} \epsilon_\K \hat n_{\K m} + \sum_{\K,m}V_{\K}^{\phantom\dagger} \left(\cre{\K m}\ddes{m} + \text{h.c.}\right)\\
&+~\epsilon\hat N + \half U \sum_{m,m'} \hat n_m \hat n_{m'}~.
\end{split}
\end{equation}
The `flavor' index $m$ takes four discrete values,  which $SU(4)$ symmetry reflects physically a combination of degenerate spin and orbital degrees of freedom: $m = (i,\sigma)$, with $i\in \{1,2\}$ denoting clockwise and anticlockwise orbits along the $z$-direction (major axis) of the CNT, and $\sigma\in\{\uparrow,\downarrow\}$ for the $z$-components of electron spin. The final two terms in \eref{eq:su4ham} represent the isolated dot, with orbital energy $\epsilon$ and charging energy $U = e^{2}/C $ ($C$ is the dot capacitance); where  $\hat n_{m}^{\phantom\dagger} = \dcre{m}\ddes{m}$, $\hat N = \sum_m \hat n_{m}^{\phantom\dagger}$ is the total dot number operator, and $\dcre{m} = \dcre{i\sigma}$ creates a $\sigma$-spin electron in orbital $i$. The first pair of terms describe the non-interacting conduction band (lead), and tunnel coupling between the dot/lead. Each is taken to be spin and orbital conserving~\cite{Choi2005,Sohnbook2001}, reflecting physically the fact (see \eg \ [\onlinecite{FinkelsteinI2007}] ) that in clean CNT devices the leads are formed within the nanotube and so `carry' the orbital symmetry, which is then conserved in the tunneling process.

The model represents the experimentally relevant situation in which the single-particle level spacing of the dot 
exceeds both the tunnel coupling to the leads and the intradot interactions.\cite{FinkelsteinII2007} In this case the four-electron `shells' of the dot are filled sequentially on sweeping the gate voltage $V_\mathrm{g}$ ($\propto -\epsilon$) and, for sufficiently-low temperatures and source-drain biases, only a single shell need be considered at a time. Only the direct Coulomb repulsion between dot electrons is moreover included; exchange interactions are generally weaker\cite{Oreg2000} and are not necessary\cite{Anders2008} to account for the experimental results of \eg\ [\onlinecite{FinkelsteinII2007}].

To the `standard model' above, we add the coupling to the external magnetic field, as well as the key SO coupling of interest here. For a field $B$ applied parallel to the nanotube axis, the Zeeman coupling to the spin and orbital degrees of freedom takes the form\cite{TieFeng2008,ourJCP}
\begin{equation}
\label{eq:hb}
\hat H_B = -B\sum_i (\gamma_{s}\hat s_i^z + \gamma_{o} \hat \tau_i^z)
\end{equation}
where 
\begin{equation}
\label{eq:siz}
\hat s_i^z = \half (\hat n_{i\uparrow}^{\pd} - \hat{n}_{i\downarrow})^{\pd}
\end{equation}
and
\begin{equation}
\label{eq:tiz}
\hat \tau_1^z = +\half (\hat n_{1\uparrow}^{\pd} + \hat{n}_{1\downarrow}^{\pd}) ~~~~~
\hat{\tau}_{2}^z = -\half (\hat{n}_{2\uparrow}^{\pd} + \hat{n}_{2\downarrow}^{\pd})
\end{equation}
are the $z$-components of the spin and orbital-pseudospin operators for orbital $i$; and where the 
coupling constants are $\gamma_s \equiv g \mu_B$ (with $g\simeq 2$ the electron $g$-factor) and $\gamma_0 \equiv 2\mu_{orb}$ (with $\mu_{orb}$ the orbital moment). 

The SO interaction obtained from detailed microscopic calculations is rather complicated\cite{Secchi2009,Ando,Heurtas,Wunsch2009}, but in practice only direct coupling between 
electron spin and the $z$-component of orbital angular momentum is relevant.\cite{Ando,Heurtas,ourJCP,TieFeng2008} 
As explained in [\onlinecite{ourJCP}], the SO interaction is then given by
\begin{subequations}
\label{eq:hso}
\begin{align}
\label{eq:hsoshorta}
\hat H_\mathrm{so} = -2&\Delta\sum_i \hat \tau_i^z \hat s_i^z \\
 =~ &\Delta ~(\hat{s}^{z}_{2} -\hat{s}_{1}^{z})
\label{eq:hsoshortb} 
\end{align}
\end{subequations}
with $\Delta$ parametrising the strength of the SO coupling; and which simple form \eg \ explains 
in essence fully the sequential tunneling experiments of [\onlinecite{rmce}].

The full Hamiltonain considered is thus 
\begin{equation}
\label{eq:fullh}
\hat{H}=\hat{H}_{SU(4)} + \hat{H}_B + \hat{H}_\mathrm{so},
\end{equation}
inclusion of $\hat{H}_B +\hat{H}_\mathrm{so}$ lowering the symmetry from $SU(4)$ to $U(1)\times U(1)\times U(1) \times U(1)$ (such that $\hat{H}$ commutes with the four charge operators $\sum_\bk\hat n_{\bk i\sigma}^{\pd} + \hat n_{i\sigma}^{\pd}$). 

Finally, to connect to experiment we simply treat $U$, $\Delta$ and the ratio $\gamma_\mathrm{o}/\gamma_\mathrm{s}$ as parameters chosen to fit experiment (which in practice is quite straightforward and unambiguous).  
The dot level energy $\epsilon$ is proportional to the experimental gate voltage, of form 
$-\epsilon = \alpha V_\mathrm{g} + \beta$; but in practice it is more convenient to work with a dimensionless gate voltage\cite{Anders2008}
\begin{equation}
\label{eq:ng}
N_\mathrm{g} = \frac{1}{2}\left(1-\frac{2\epsilon}{U}\right)
\end{equation}
such that differences in $V_\mathrm{g}$ are proportional to differences in $N_\mathrm{g}$ (with a proportionality constant determined if desired by fit to experiment).


\section{Theoretical background}
\label{sec:theory}
 The dot Green function $G_{i\sigma}(\omega) \leftrightarrow G_{i\sigma}(t) =-i \theta(t) \langle \{ c_{i\sigma}(t), c^\dagger_{i\sigma}\} \rangle$ is central to understanding transport through the dot; which is directly related to the single-particle spectrum $D_{i\sigma}(\omega) = - \tfrac{1}{\pi}\Im G_{i\sigma}(\omega)$ \via\ the Meir-Wingreen formula~\cite{meirwingreen}, as now briefly summarised.

One first partitions the conduction band of \eref{eq:su4ham} into two equivalent leads, left ($L$) and right ($R$). These are taken conventionally~\cite{hewsonbook}  to be flat bands of width $2D$, with density of states $\rho(\omega) = \rho_0 = 1/(2D)$ for $|\omega| < D$ (and with $D$ by far the largest energy scale in the model). The tunneling matrix elements to the $L,R$ leads are taken  for simplicity as $V_L$ and $V_R$, independent of $\K$. The leads are fixed at different chemical potentials, $\mu_L$ and $\mu_R$, with a bias voltage $V_\mathrm{sd}$ between them, $\mu_L - \mu_R = e V_\mathrm{sd}$. After transients have subsided, the bias voltage gives rise to a steady state current through the dot, $J$, carried by its four conduction channels.  An exact expression for $J$
follows from the Keldysh formalism:\cite{meirwingreen}
\begin{equation}
\label{eq:j}
J = \frac{e}{h} G_0 \pi (\Gamma_L + \Gamma_R) \int_{-\infty}^\infty \D\omega\left[f_L(\omega)-f_R(\omega)\right] \sum_{i,\sigma} D_{i\sigma}(\omega)
\end{equation}
where
\begin{equation}
\label{eq:gnought}
G_0 = \frac{4\Gamma_L\Gamma_R}{(\Gamma_L+\Gamma_R)^2},
\end{equation}
$\Gamma_\nu = \pi|V_\nu|^2\rho_0$ is the hybridization strength of the dot to lead $\nu$ ($=L$ or $R$), and $f_\nu(\omega) = \left[e^{\beta(\omega-\mu_\nu)}+1)\right]^{-1}$ is the Fermi function for the lead with inverse temperature $\beta = 1/T$ ($k_\mathrm{B} =1$).  It is convenient to define $\Gamma_L+\Gamma_R = \Gamma$ (we later take $\Gamma$ to be the `unit' of energy), such that the relative strength of coupling to the $L$ and $R$ leads enter through the dimensionless $G_0$ (which can be chosen according to the experimental setup under consideration). 
In the perfectly symmetric case of $\Gamma_L = \Gamma_R$, $G_0 =1$ is maximal, while in the extreme asymmetric case of $\Gamma_L \ll \Gamma_R$ (say), $G_0\sim 4\Gamma_L/\Gamma_R \ll 1$. 

The key experimental quantity is the differential conductance, $ G_\mathrm{c}(V_\mathrm{sd}) = \D J/\D V_\mathrm{sd}$. In the zero-bias limit, gives an exact expression for $G_\mathrm{c}(0)\equiv G_\mathrm{c}^0$ 
in terms of the equilibrium single-particle spectrum:
\begin{equation}
\label{eq:gczero}
G_\mathrm{c}^0 = \frac{e^2}{h} G_0 \pi \Gamma \int_{-\infty}^{\infty} \D\omega \left(-\frac{\partial f}{\partial\omega}\right)\sum_{i,\sigma} D_{i\sigma}(\omega)
\end{equation}
with $f(\omega)=(e^{\beta\omega}+1)^{-1}$, which further reduces to
\begin{equation}
\label{eq:gczerozerot}
G_\mathrm{c}^0 \overset{T\to 0}= \frac{e^2}{h} G_0 \pi \Gamma \sum_{i,\sigma} D_{i\sigma}(0)
\end{equation}
for $T=0$. $D_{i\sigma}(\omega)$ can be calculated accurately at equilibrium using the recent FDM NRG 
method~\cite{PetersPruschkeAnders2006,WeichselbaumDelft2007}.

To make connection to experiments at finite source-drain bias, note that while \eref{eq:j} itself remains exact for finite $V_{\mathrm{sd}}$, the difficulty lies in calculating $D_{i\sigma}(\omega)$ out of equilibrium.
While recent progress has been made in applying NRG to the non-equilibrium single-impurity Anderson model (see \eg \ [\onlinecite{AndersNoneq2008}]) it is currently prohibitive to apply these techniques to the model of \eref{eq:fullh}. As in previous work~\cite{Anders2008,DEL2Level} we thus make the standard approximation of neglecting the $V_\mathrm{sd}$-dependence of the dot self-energy. Using \eref{eq:j} this leads to
\begin{multline}
\label{eq:gencond}
G_\mathrm{c}(V_\mathrm{sd}) =\\ \frac{e^2}{h} G_0 \pi\Gamma \int_{-\infty}^{\infty}\frac{\D\omega}{2}\left(-\frac{\partial f_L(\omega)}{\partial\omega}-\frac{\partial f_R(\omega)}{\partial\omega}\right)\sum_{i,\sigma}D_{i\sigma}(\omega)
\end{multline}
where we have taken a symmetric voltage split \cite{DEL2Level} between the leads, $\mu_{L/R} = \pm eV_\mathrm{sd}/2$. Equations (\ref{eq:gczero})--(\ref{eq:gencond}) form the basis of our calculations of \sref{sec:results}.

\subsection{Friedel sum rule}
\label{sec:fsr}
Eq.~(\ref{eq:gczerozerot}) relates exactly the zero-bias conductance at $T=0$ to the four spectra $D_{i\sigma}$ at the Fermi level, $\omega=0$. These in turn can be obtained exactly in terms of the so-called `excess charges' of the dot in the four distinct conduction channels, as now sketched.

The Green function $G_{i\sigma}(\omega)$ is diagonal in spin and orbital indices, and given by
\begin{equation}
\label{eq:dyson}
G_{i\sigma}(\omega) = \left[\omega + \I0^+ -\epsilon_{i\sigma} - \Gamma(\omega) - \Sigma_{i\sigma}(\omega)\right]^{-1}
\end{equation}
where $\Gamma(\omega) = \Gamma \left[\pi^{-1}\ln|(\omega+D)/(\omega-D)| - \I\theta(D-|\omega|)\right]$ is the $\omega$-dependent hybridization function, $\epsilon_{i\sigma}$ the effective one-electron energy under $\hat{H}$ (\eref{eq:fullh}), and $\Sigma_{i\sigma}(\omega)$ is the dot self-energy. Luttinger's integral theorem, \cite{luttward,hewsonbook}
\begin{equation}
\label{eq:lutt}
\Im \int_{-\infty}^0 \D\omega \left(\frac{\partial \Sigma_{i\sigma}(\omega)}{\partial \omega}\right)G_{i\sigma}(\omega) = 0,
\end{equation}
applies separately within \emph{each} conduction channel $(i,\sigma)$, allowing one to follow the steps of \eg\ [\onlinecite{hewsonbook}] to derive the Friedel sum rule:~\cite{hewsonbook,Langreth1966}
\begin{equation}
\label{eq:fsr}
\delta_{i\sigma} = \pi n_{\mathrm{imp};i\sigma}^{}
\end{equation}  
This relates $\delta_{i\sigma}$, the (Fermi level) phase shift of the conduction electrons in the $(i,\sigma)$ channel, to the corresponding excess charge given by
\begin{equation}
\label{eq:nimpdef}
n_{\mathrm{imp};i\sigma}^{} = \int_{-\infty}^0 \D\omega \left\{D_{i\sigma}(\omega) + \sum_{\K} \left[D_{\K i\sigma}(\omega) - D_{\K i\sigma}^0(\omega)\right]\right\}
\end{equation}
where $D_{\K i\sigma}(\omega)$ [$D_{\K i\sigma}^0(\omega)$] is the $(i,\sigma)$ conduction electron spectrum for wavevector $\K$ in the presence [absence] of the dot. The Fermi level value of the spectrum at $T=0$ is readily shown to satisfy \cite{hewsonbook}
\begin{equation}
\label{eq:sinsquared}
\pi\Gamma D_{i\sigma}(0) = \sin^2(\delta_{i\sigma}),
\end{equation}
and hence from \eref{eq:gczerozerot} we obtain
\begin{equation}
\label{eq:finalfsr}
G_\mathrm{c}^0 \overset{T\to 0}= \frac{e^2}{h} G_0 \sum_{i,\sigma} \sin^2(\pi n_{\mathrm{imp};i\sigma}^{}).
\end{equation}
The $T=0$ zero-bias conductance is thus related to the excess charges in the four conduction channels (themselves readily obtained via a thermodynamic NRG calculation). In the experimentally relevant limit where $D$ is the largest energy scale, these excess charges are moreover confined to the dot itself. One can then approximate $n_{\mathrm{imp};i\sigma}^{}$ by $\langle \hat{n}_{i\sigma}^{}\rangle$, thereby producing a simple relationship between the dot occupancy and its transport properties.

In the $SU(4)$-symmetric limit ($\Delta = 0 = B$) considered previously,\cite{Anders2008} the excess charges are equivalent in all four channels and hence $n_{\mathrm{imp};i\sigma}^{} = n_\mathrm{imp}^{}/4$ with $n_\mathrm{imp}^{} = \sum_{i,\sigma} n_{\mathrm{imp};i\sigma}^{}$. Eq.~(\ref{eq:finalfsr}) then reduces to\cite{Anders2008}
\begin{equation}
\label{eq:fsrsu4}
G_\mathrm{c}^0 \overset{T\to 0}= \frac{4e^2}{h} G_0 \sin^2\left(\frac{\pi n_\mathrm{imp}^{}}{4}\right).
\end{equation}

\subsection{NRG}
\label{sec:nrg}
We analyse the model \eref{eq:fullh} using the numerical renormalization group (NRG). This  technique~\cite{NRGrmp} has long provided access to numerically-exact results for thermodynamics and, with the recent identification of its complete Fock space,\cite{AndersSchillerBasisPRB} an equally systematic and reliable route to dynamical properties.

The basic approach is detailed in \eg\ [\onlinecite{kww1}]. A logarithmic discretization of the conduction band states is first used to map the Hamiltonian onto a countably infinite one-dimensional chain. The linear chain is diagonalised iteratively, starting from a single site and adding the others one by one. The key advantage of  logarithmic discretization is that the coupling constants along the chain decrease rapidly, and 
the high-energy states of one iteration can be discarded without affecting the low-energy states  retained in later iterations. As such, a fixed number of states can in practice be kept at each iteration, rendering the iterative diagonalization of the Hamiltonian numerically tractable.

The information obtained from each iteration allows one to build up the thermodynamics and dynamics of the model. Eigenstates of a given iteration are used to calculate thermodynamics at an appropriately-chosen temperature (sufficiently low that discarded states of earlier iterations are unimportant, yet sufficiently high that the energy splittings of later iterations are thermally smeared out). This effective temperature decreases exponentially with the iteration number, and hence allows access to thermodynamics on all physical energy scales after only a modest number of iterations.

Dynamics of the model are calculated by means of the recent observation \cite{AndersSchillerBasisPRL,AndersSchillerBasisPRB} that the set of all \emph{discarded} states forms a complete basis of the discretized NRG Hamiltonian. By expanding the full density matrix in this basis, accurate results for dynamical correlation functions may be obtained over a wide range of frequency and temperature scales.\cite{PetersPruschkeAnders2006,WeichselbaumDelft2007} Dynamics of the discrete NRG Hamiltonian necessarily arise as a series of isolated poles: in order to capture the behavior of the original continuous model,one then convolves the discrete spectra with an appropriate broadening function on a logarithmic scale (see \eg\ [\onlinecite{NRGrmp}]). Potential artefacts of the discretization process are minimised in three standard ways: dot-lead couplings are premultiplied by the standard $A_\Lambda$ factor,\cite{kww1,CampoOliveira2005} the Oliveira `$z$-averaging'\cite{Oliveira1994} is used to average discrete spectra with different logarithmic discretizations, and the Green function is obtained not directly but from its self-energy \cite{bullahewprus}, calculated as a ratio of two correlation functions where any remaining discretization effects largely cancel.  

The calculations in this work have been obtained with an NRG discretization parameter $\Lambda = 3$, exploiting the full $U(1)\times U(1)\times U(1)\times U(1)$ symmetry of the model. We have typically averaged results for five different $z$s and have kept the lowest $2500$--$4500$ states at each iteration.


\section{Atomic limit and physical picture}
\label{sec:atomic}
Here we show that a physical understanding of the interplay between Kondo physics, SO and Zeeman couplings, follows readily by considering the isolated dot (the atomic limit, $\Gamma =0$).The latter has been considered 
in [\onlinecite{ourJCP}], from which we take results as required; denoting the dot states in an obvious notation as $|\uparrow;-\rangle = \dcre{1\uparrow}|-;-\rangle$, $|-;\downarrow\rangle = \dcre{2\downarrow}|-;-\rangle$, $|-;\uparrow\downarrow\rangle = \dcre{2\uparrow}\dcre{2\downarrow}|-;-\rangle$, and so on.

\begin{figure}
\includegraphics{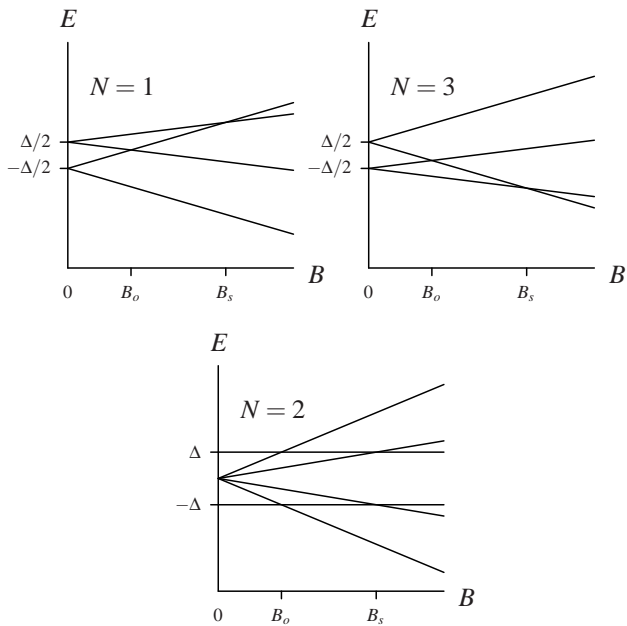}
\caption{\label{fig:alenergies} Splittings of the $N=1,2,3$
atomic limit states, for finite $B$ and $\Delta >0$ (relative to their values for $B=0=\Delta$).}
\end{figure}

The energies of the 16 possible isolated dot states follow directly from \eref{eq:fullh} with $V_\bk=0$, and may be classified by their total occupation number $N = \langle \hat N \rangle$. In the absence of SO and Zeeman couplings, $\Delta =0 = B$, the $4!/[N!(4-N)!]$ $N$-electron states are degenerate, with energies
${\cal{E}} = N\epsilon + N(N-1)U/2$. On switching on $\Delta$ and $B$, the $N=0$ and $N=4$ states are trivially unaffected, while the $N=1 -3$ electron states are split as shown schematically in \fref{fig:alenergies}. 
For $B=0$, SO coupling splits both the $N=1$ and $N=3$ states into two degenerate pairs separated by an energy $\Delta$. The $N=2$ states by contrast are split into three groups, of degeneracies $1$, $4$ and $1$, with relative energies $-\Delta$, $0$ and $\Delta$ respectively.

Switching on the field further splits the states by both spin- and orbital-Zeeman effects (\fref{fig:alenergies}), and in the $N=1$ sector a singly-degenerate ground state ($|\uparrow;-\rangle$) arises for all $B>0$.
The situation is somewhat more complicated in the $N=2$ and $N=3$ sectors, since in both cases competition between SO and Zeeman effects leads to level crossings in the ground state. In the $N=2$ sector it is the \emph{orbital} Zeeman effect that competes with SO coupling: the low-field ground state is  $|\uparrow;\downarrow\rangle$ as favored by the SO interaction \eref{eq:hsoshortb}, while at higher fields the ground state $|\uparrow\downarrow;-\rangle$ is favored by the orbital Zeeman interaction (\eref{eq:hb}) for the experimentally relevant case $\gamma_{\mathrm{o}} >\gamma_{\mathrm{s}}$; the ground state level crossing occuring at a field
\begin{equation}
\label{eq:bodef}
B_\mathrm{o}=\frac{\Delta}{\gamma_\mathrm{o}}.
\end{equation}
In the $N=3$ sector by contrast it is \emph{spin} Zeeman which now competes with SO coupling, producing a level crossing from the low-field ground state $|\uparrow\downarrow;\downarrow\rangle$ to $|\uparrow\downarrow;\uparrow\rangle$ at a field $B_{\mathrm{s}}$ ($>B_{\mathrm{o}}$) given by
\begin{equation}
\label{eq:bsdef}
B_\mathrm{s}=\frac{\Delta}{\gamma_\mathrm{s}}.
\end{equation}
These special values of the field turn out to be central to the Kondo physics of the model, as explained below.

\begin{figure}
\includegraphics[scale=0.9]{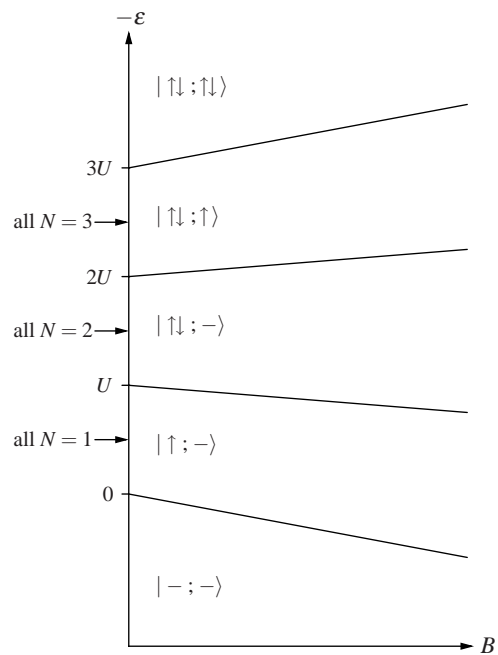}
\caption{\label{fig:cbalnoso} Schematic atomic limit ($\Gamma=0$)  `phase diagram' in the absence of SO coupling, as a function of $B$ and $-\epsilon\propto V_\mathrm{g}$, showing boundaries between ground states of different charge.}
\end{figure}

From the atomic limit energies, the ground state `phase diagram' is readily constructed as a function of $B$ and $-\epsilon$ ($\propto  V_g$). \Fref{fig:cbalnoso} first shows the situation where SO coupling is absent, $\Delta=0$; solid lines
marking the boundaries between states of different ground state charge. For any fixed $B$, increasing $-\epsilon$ generates the familiar Coulomb blockade (CB) staircase. When $\epsilon$ lies sufficiently in excess of the Fermi level, both orbitals are empty; and on lowering $\epsilon$ through the Fermi level the total number of electrons on the dot increases stepwise from zero to four. 
Notice that, for all $B$, \fref{fig:cbalnoso} is symmetric under reflection about the line $-\epsilon=\tfrac{3}{2}U$
corresponding to the midpoint of the $N=2$-electron valley (\ie \ to replacing $\epsilon \rightarrow -[\epsilon +3U]$); reflecting for $\Delta =0$ the essential equivalence of states with $N$ and $4-N$ electrons under a particle-hole transformation (specifically $d_{1\sigma}^{\dagger} \leftrightarrow d^{\pd}_{2-\sigma}$).
And at points of degeneracy between $N$ and $N+1$ electrons (\fref{fig:cbalnoso} solid lines), there is naturally facile zero-bias sequential tunneling through the dot (and hence enhanced conductance) when it is connected to the leads.\cite{ourJCP}

 Degeneracies between states of the \emph{same} total charge do not promote sequential tunneling, but are of course vital for the Kondo effects arising from coherent cotunneling processes on coupling to the leads (discussed below). Such degeneracies arise at zero-field in the $N=1$, $N=2$ and $N=3$ valleys, where all states of given $N$ are degenerate. For any $B >0$ however, there is an immediate `transition' to a singly-degenerate state in each case with maximal $\tau^z =\tfrac{1}{2}(n_{1}-n_{2})$.

\begin{figure}
\includegraphics[scale=0.9]{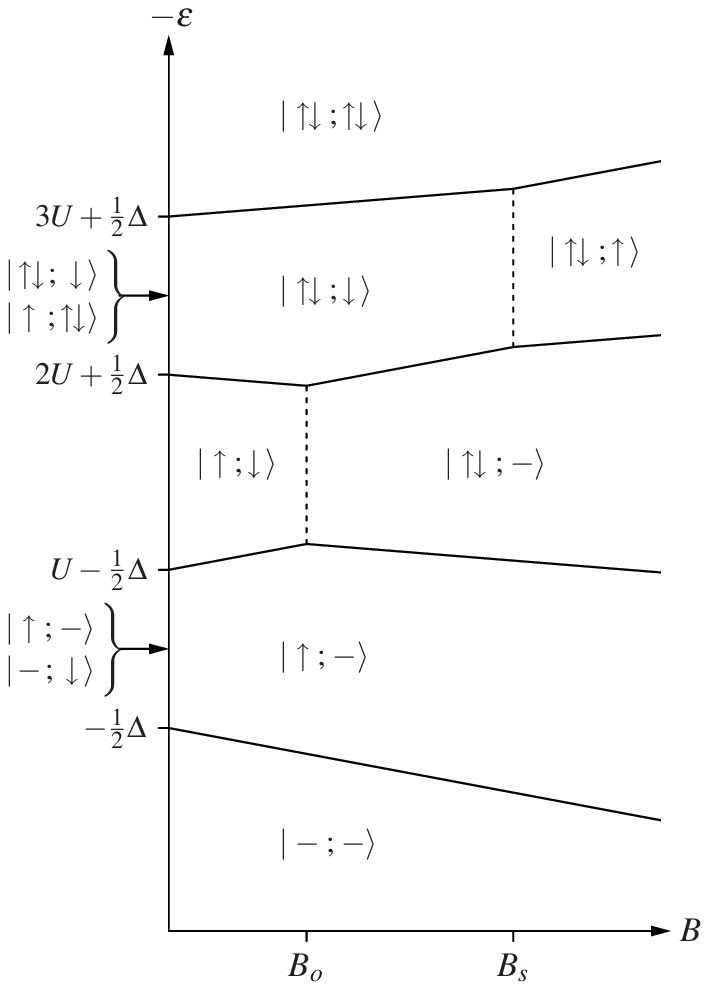}
\caption{\label{fig:cbal} As \fref{fig:cbalnoso}, but for the case of a finite SO coupling $\Delta$. Boundaries between ground states with the same charge are shown by dashed lines.}
\end{figure}

  The above situation changes qualitatively on introduction of SO coupling; as illustrated in \fref{fig:cbal},
showing the generic behavior for $\Delta>0$ and $\gamma_o > \gamma_s$ (albeit for illustration using a somewhat smaller ratio  $\gamma_o/\gamma_s = B_{\mathrm{s}}/B_{\mathrm{o}}$ than seen experimentally). 
This structure is clearly more interesting than  \fref{fig:cbalnoso}. First, the SO coupling $\Delta$ has a dramatic effect at zero field. In both the $N=1$ and $N=3$ valleys the degeneracy of the ground state at $B=0$ is reduced from 4 to 2 (as in \frefs{fig:alenergies}{fig:cbal}); the degree of freedom associated with this two-fold degeneracy being neither a pure spin nor an orbital pseudospin, but a mixture of the two. In the $N=2$ valley the effect of $\Delta$ is even more severe: the 6-fold degenerate ground state for $\Delta=0$ is replaced by the non-degenerate state $ |\uparrow;\downarrow\rangle$, the other five states again lying $\mathcal{O}(\Delta)$ higher (\cf\ \fref{fig:alenergies}).

The finite-field ground state level crossings in the $N=2, 3$ electron sectors (\fref{fig:alenergies}), mean as shown in \fref{fig:cbal} (dashed lines) that for all $\epsilon$ when $N=2$ there is a crossing from $|\uparrow;\downarrow\rangle$ to $|\uparrow\downarrow;-\rangle$ at $B=B_\mathrm{o}$\cite{TieFeng2008};
and likewise throughout the $N=3$ sector, a crossing from $|\uparrow\downarrow;\downarrow\rangle$ to $|\uparrow\downarrow;\uparrow\rangle$ at $B=B_{\mathrm{s}}$. Associated with these ground state crossovers are 
naturally kinks in the CB steps seen in \fref{fig:cbal} (solid lines), as discussed in [\onlinecite{ourJCP}] and observed in the sequential tunneling experiments of [\onlinecite{rmce}] on ultraclean CNT dots.
By contrast there is no ground state level crossing in the $N =1$ sector (\fref{fig:alenergies}). Hence, 
as evident in \fref{fig:cbal}, the `reflection symmetry' seen in \fref{fig:cbalnoso} for $\Delta =0$ is absent -- states with $N$ and $4-N$ electrons no longer being equivalent under the particle-hole transformation $d_{1\sigma}^{\dagger} \leftrightarrow d^{\pd}_{2-\sigma}$ (arising because
$\hat{H}_{\mathrm{so}}(\Delta) \rightarrow \hat{H}_{\mathrm{so}}(-\Delta)$ under such).

As mentioned above the significance of degeneracies between same-charge states is that, on coupling to the leads, their associated low energy degrees of freedom can be screened by many-body Kondo effects which enhance 
conductance through the dot. \cite{nglee,glazmanraikh} Two basic Kondo effects may in fact arise,\cite{Lim2006}
$SU(4)$ and $SU(2)$. Only when the full model is close to being $SU(4)$-symmetric does the former arise (we define `close to' shortly); the more common case, occuring for doubly-degenerate atomic limit states where the low-energy effective model maps onto a spin-$\half$ Kondo model, is the $SU(2)$ Kondo effect. One key physical distinction between the two is the low-energy/temperature Kondo scale on which they are manifest, generically denoted $T_\mathrm{K}$. In the strongly correlated regime $U\gg \Gamma$, the Kondo scales in the two cases are \cite{hewsonbook}
\begin{subequations}
\label{eq:kondoscales}
\begin{align}
T_{\mathrm{K}}^{SU(2)} &\sim \Gamma \exp\left(-\frac{\pi U}{8\Gamma}\right)\\
T_{\mathrm{K}}^{SU(4)} &\sim \Gamma \exp\left(-\frac{\pi U}{16\Gamma}\right),
\end{align}
\end{subequations}
modulo prefactors that depend weakly on the bare parameters, such that $T_\mathrm{K}^{SU(4)} \gg T_\mathrm{K}^{SU(2)}$.

With the above in mind, the essential qualitative physics of the model is readily deduced.
We start with $\Delta=0$, and consider specifically the zero-temperature limit (the effect of temperature being
simply to smear out the pristine $T=0$ behavior over an energy scale $\mathcal{O}(T)$). The appropriate atomic limit picture for $\Delta=0$ is \fref{fig:cbalnoso}. On coupling to the leads the CB steps, and associated zero-bias conductance arising from facile sequential tunneling, are broadened over an energy scale $\mathcal{O}(\Gamma)$.\cite{ourJCP} In the $N=1$, $2$ and $3$ electron valleys for $B=0$, $SU(4)$ Kondo effects take place\cite{Choi2005,Mitchell2006,Busser2007,Anders2008} (effective low-energy $SU(4)$ Kondo, or Coqblin-Schrieffer, models~\cite{hewsonbook}  being obtained \via \ a Schrieffer-Wolff transformation, on retaining the $4!/[N!(4-N)!]$ degenerate dot states in the ground state manifold and the cotunneling processes that connect them). The $SU(4)$ Kondo physics in the $N=1-3$ valleys will naturally persist at finite fields until $\gamma_\mathrm{s} B \sim {\cal O}(T_\mathrm{K}^{SU(4)})$; while for field strengths in excess of the $SU(4)$ Kondo scale, the Kondo effect will be destroyed and the conductance will be correspondingly low.
And for $\Delta =0$, the conductance in the $N=1$ and $N=3$ electron valleys as a function of $B$ (or $T$) will be coincident; reflecting the equivalence of associated states under the particle-hole transformation discussed above.

On introducing a finite $\Delta$, the key quantity in determining whether $SU(4)$ or $SU(2)$ Kondo physics prevails
is the ratio $\xi =\Delta/T_\mathrm{K}^{SU(4)}$ (with $T_\mathrm{K}^{SU(4)}$ the $SU(4)$ scale in the absence of SO coupling). For $\xi \ll 1$ the zero-field $SU(4)$ Kondo effects described above are still favorable, since the Kondo stabilization energy outweighs the splittings of the atomic limit states in \fref{fig:alenergies}. For these small $\Delta$s, one  thus expects the physics to be essentially unchanged from the $\Delta=0$ limit. Only when $\Delta$ becomes comparable to $T_\mathrm{K}^{SU(4)}$  will it have a noticeable effect.

For $\Delta \gg T_\mathrm{K}^{SU(4)}$ by contrast the appropriate starting picture is now \fref{fig:cbal}, and $SU(4)$ Kondo effects no longer arise. The $N=2$ valley will not exhibit any Kondo effect at zero-field, since its ground state is singly degenerate.  The $N=1$ and $N=3$ valleys for $B=0$ will however display $SU(2)$ Kondo effects involving their doubly-degenerate ground states (\fref{fig:cbal}): an effective low-energy Kondo model of $SU(2)$ form obviously arises in each case, under a Schrieffer-Wolff transformation retaining the appropriate degenerate pair of dot states indicated in \fref{fig:cbal}.

For $B>0$, the zero-field Kondo effects in the $N=1,3$ valleys will again `spill over' into the $B$-plane by an amount of order their $SU(2)$ Kondo scales (rather less than in the $SU(4)$ case, as above). 
In addition however, the level crossings occurring at finite fields in both the $N=2$~~\cite{TieFeng2008} and $N=3$ valleys (\fref{fig:cbal}) means that additional $SU(2)$ Kondo effects will now occur at fields $B=B_\mathrm{o}$ and $B=B_\mathrm{s}$ respectively (the two-fold ground state degeneracy of the free dot states at either field 
generating an $SU(2)$ Kondo model under Schrieffer-Wolff). These will be discernable as long as they are not subsumed by the zero-field $SU(2)$ Kondo effect. This is clearly not an issue in the $N=2$ valley, no zero-field Kondo effect occurring here anyway for $\Delta \gg T_\mathrm{K}^{SU(4)}$; 
while in the $N=3$ valley it requires $(\Delta \equiv $) $\gamma_\mathrm{s} B_\mathrm{s}  \gg  T_\mathrm{K}^{SU(2)}$, readily seen to be satisfied since $T_\mathrm{K}^{SU(2)}\ll T_\mathrm{K}^{SU(4)}$. In the $N=1$ valley by contrast, the absence of a ground state level crossing at finite field (\fref{fig:cbal}) means that the $SU(2)$ Kondo effect arising here at zero-field will simply be steadily destroyed with increasing $B$, dying out on a scale of order $\gamma_{\mathrm{s}}B \sim {\cal{O}}(T_{\mathrm{K}}^{SU(2)})$. And since the $N=1$ and $N=3$ electron valleys in particular exhibit distinct behavior as a function of field, then, as for the atomic limit states themselves, the $N \leftrightarrow 4-N$ symmetry of the conductance as a function of $-\epsilon \propto V_{\mathrm{g}}$ is again absent for $\Delta \neq 0$.


\section{Results}
\label{sec:results}
The above considerations are purely qualitative, and we have analysed the model in detail \via \ NRG,
over a large parameter space. Here we present a selection of results, focusing in particular on parameter regimes 
applicable to experiment. Specifically, we make comparison to two experimental works: Makarovski \etal\ \cite{FinkelsteinI2007} (denoted `M') and Jarillo-Herrero \etal\ \cite{jarillo2} (`JH'). The former device is somewhat more strongly correlated than the latter (although in both cases the ratio $U/\Gamma$ is sufficiently large to generate non-trivial Kondo behavior), so we can compare theory to experiment in two distinct physical regimes. Details of how the model parameters are chosen will be given at appropriate points in the following. It will also be convenient to define and use the reduced parameters: $\tilde U=U/\Gamma$, $\tilde \Delta = \Delta/\Gamma$, $\tilde T = T/\Gamma$ and $\tilde B = \gamma_\mathrm{s}B/\Gamma$.

\subsection{Zero-bias conductance at $B=0$}
\label{sec:zbb0}
\begin{figure}
\includegraphics[scale=0.9]{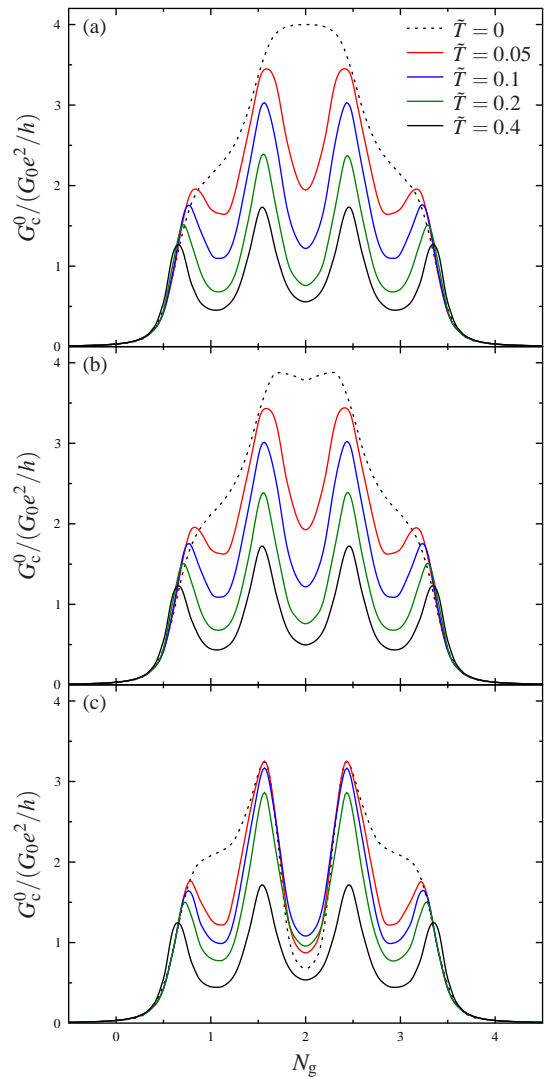}
\caption{\label{fig:zbcondv}(Color online.) Zero-bias conductance $G_\mathrm{c}^{0}/(G_0 e^2/h)$ as a function of dimensionless gate voltage $N_\mathrm{g}$, at $B=0$ and for $\tilde U=20$. Each panel shows the conductance for the range of temperatures indicated, with SO coupling strengths (a) $\tilde\Delta = 0$, (b) $\tilde\Delta = 0.02 \simeq T_{\mathrm{K}}^{SU(4)}$ and (c) $\tilde\Delta =0.2 \gg T_{\mathrm{K}}^{SU(4)}$.}
\end{figure}

\Fref{fig:zbcondv} shows the $B=0$ zero-bias conductance as a function of the dimensionless gate voltage
$N_{\mathrm{g}} =\tfrac{1}{2}(1-\tfrac{2\epsilon}{U})$ (\sref{sec:model}), for a range of temperatures $(T)$ and three different SO coupling strengths. Here we take an interaction 
$\tilde U = U/\Gamma=20$, so that the relative widths of the CB peaks for $T\gg T_\mathrm{K}$ are in line with the experiments of M (\cf \ the discussion in [\onlinecite{Anders2008}]). 

\Fref{fig:zbcondv}(a) shows the $\Delta=0$ case, considered in  [\onlinecite{Anders2008}].
The $T=0$ conductance (dotted line) evolves stepwise with $N_\mathrm{g}$, and follows \eref{eq:fsrsu4} as a function of $n_\mathrm{imp}^{}$; with a stepwise increase in $n_\mathrm{imp}^{}$  itself  as the dimensionless gate voltage is increased.\cite{Anders2008} The latter is a result of the relatively large $U/\Gamma$, leading to strong charge quantization on the dot except when $\epsilon$ is within $\mathcal{O}(\Gamma)$ of the atomic limit charge-degeneracy points at $N_\mathrm{g} = \tfrac{1}{2}$, $\tfrac{3}{2}$, $\tfrac{5}{2}, \tfrac{7}{2}$. On increasing $T$ the zero-bias conductance is rapidly eroded towards the centers of the $N_\mathrm{g}=1$, $2$ and $3$ electron valleys, resulting in the familiar Coulomb-blockade valley structure with conductance peaks around the atomic limit charge-degeneracy points. The temperature scale over which the erosion takes place is of course $T_\mathrm{K}^{SU(4)}$, given by \eref{eq:kondoscales} and obtained numerically~\cite{footnoteTK}
as $T_\mathrm{K}^{SU(4)}\simeq 0.02\Gamma$ [$0.03\Gamma$] for $N_\mathrm{g} = 2$ [$1$]. For $T_\mathrm{K}^{SU(4)}  \ll T \ll \Gamma$, the HWHM of the Coulomb blockade peaks are of order 
$\mathcal{O}(\Gamma)$ (not \emph{precisely} $\Gamma$ due\cite{Anders2008} to electron interactions); while for $T\gg \Gamma$ the CB peaks simply broaden to become of width $\mathcal{O}(T)$ instead. 

\Fref{fig:zbcondv}(b) shows the effect of a non-zero SO coupling of order $\Delta \sim T_\mathrm{K}^{SU(4)}$, such that SO coupling competes with the $SU(4)$ Kondo effects in the centers of the Coulomb blockade valleys. The $T=0$
conductance is slightly eroded around $N_\mathrm{g}\sim 2$, but is qualitatively unchanged elsewhere. Upon increasing $T$, it is clear that once $T\gtrsim\Delta$, the conductance appears essentially identical to the $\Delta=0$ case \fref{fig:zbcondv}(a), as one expects physically.

On now considering $\Delta \gg T_\mathrm{K}^{SU(4)}$ (but still small relative to the non-universal scale $\Gamma$), the situation changes to that of \fref{fig:zbcondv}(c). Here the Kondo effect in the two-electron Coulomb-blockade valley at $N_\mathrm{g}\sim 2$ is destroyed at $T=0$ as expected (\sref{sec:atomic}), and the conductance remains rather low for all temperatures shown.  For the $N_{\mathrm{g}}=1$ and $3$ electron CB valleys by contrast, the $T=0$ conductance is still $G_{\mathrm{c}}^{0}/G_{0}\simeq 2e^2/h$ (as for the $SU(4)$ symmetric limit $\Delta =0$, 
which follows in that case from \eref{eq:fsrsu4} with $n_{\mathrm{imp}} \simeq 1$ $(3)$ for the center of the $N_{\mathrm{g}}=1$ $(3)$ electron CB valley). This behavior for \emph{large} $\Delta $ is now however symptomatic of the $SU(2)$ Kondo effect arising in that case; as readily understood using the general result \eref{eq:finalfsr} for $G_{\mathrm{c}}^{0}$, considering the $N_{\mathrm{g}}=1$ case explicitly. Recall (\fref{fig:cbal}) that the free dot ground state for $N_{\mathrm{g}}=1$ is the degenerate pair $|\uparrow;-\rangle$ and $|-;\downarrow\rangle$ (which generate $SU(2)$ Kondo under Schrieffer-Wolff on cotunneling to the leads, \sref{sec:atomic}), 
for which $\langle\hat{n}_{1\uparrow}\rangle =\tfrac{1}{2} =\langle\hat{n}_{2\downarrow}\rangle$ and $\langle\hat{n}_{1\downarrow}\rangle =0=\langle\hat{n}_{2\uparrow}\rangle$; and since $n_{\mathrm{imp};i\sigma}^{} \simeq \langle\hat{n}_{i\sigma}\rangle$ as noted in \sref{sec:fsr}, \eref{eq:finalfsr} gives directly
$G_{\mathrm{c}}^{0}/G_{0} \simeq 2e^{2}/h$. Note however that although the $T=0$ conductance for $N_{\mathrm{g}}=1$ or $3$ barely discriminates between $SU(4)$ ($\Delta =0$, \fref{fig:zbcondv}(a)) and $SU(2)$ ($\Delta \gg T_{\mathrm{K}}^{SU(4)}$, \fref{fig:zbcondv}(c)), the erosion of conductance with temperature occurs more rapidly in the latter case -- occurring naturally on the $SU(2)$ Kondo scale $T_{\mathrm{K}}^{SU(2)}$ (with 
$T_{\mathrm{K}}^{SU(2)}/\Gamma \simeq 8\times 10^{-3}$ in \fref{fig:zbcondv}(c)).

\begin{figure}
\includegraphics{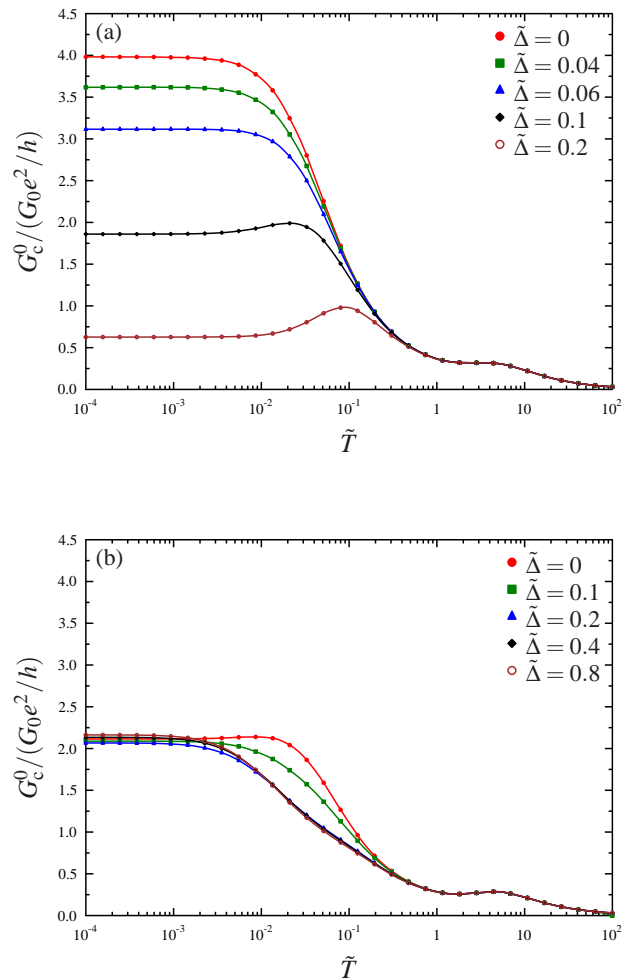}
\caption{\label{fig:zbcondt}(Color online.) Zero bias conductance $G_\mathrm{c}^{0}/(G_0 e^2/h)$ \emph{vs}
$\tilde T$ for various SO couplings $\tilde\Delta$ as indicated, at $B=0$ with $\tilde U=20$. For (a) the center of the $N_{\mathrm{g}}=2$-electron CB valley; (b) the center of the $N_{\mathrm{g}}=1$ valley.
For $B=0$, the behavior for $N_\mathrm{g}=3$ is identical to (b) by symmetry (see \fref{fig:alenergies}).}
\end{figure}

The full $T$-dependence of the conductance in the centers of the CB valleys is shown in \fref{fig:zbcondt},
$G_{\mathrm{c}}^{0}$ \emph{vs} $\tilde T$ (on a log-scale) for a range of $\tilde\Delta$ values, and for $N_{\mathrm{g}}=2$ (\fref{fig:zbcondt}(a)) and $N_{\mathrm{g}}=1$ (\fref{fig:zbcondt}(b), $N_{\mathrm{g}}=1$ and $3$ being equivalent by symmetry for $B=0$, see \eg \ \fref{fig:alenergies}).
Results for $\tilde\Delta=0$ have been considered in [\onlinecite{Anders2008}]: in each valley the strong conductance enhancement due to coherent $SU(4)$ Kondo transport is evident in the `Kondo plateau' for temperatures
$T\lesssim T_\mathrm{K} \equiv T_{\mathrm{K}}^{SU(4)}$ (although note that the universal scaling forms of 
$G_\mathrm{c}^{0}(T)/G_\mathrm{c}^{0}(T$$=0)$ in the two valleys differ quantitatively,\cite{Anders2008} reflecting the two distinct $SU(4)$ Kondo effects that arise therein). On a temperature scale $T\sim U/2$, a conductance shoulder is also evident, corresponding to incoherent sequential tunneling transport.

In the $N_{\mathrm{g}}=2$-electron valley (\fref{fig:zbcondt}(a)) the Kondo plateau is progressively destroyed 
on increasing the SO coupling $\tilde\Delta$, as all but the lowest (non-degenerate) atomic limit states become projected out of the low-energy manifold (see \fref{fig:alenergies}). For $\Delta$ sufficiently large compared to $T_K^{SU(4)}$, a peak is seen to emerge in the conductance at $T \sim {\cal{O}}(\Delta)$ (in practice $T\simeq  \Delta/2$) -- naturally  so, this being the energy gap to higher SO-split states (\fref{fig:alenergies}), and mixing in of which enhances the conductance. And in all cases shown, the high-temperature ($T\gg\Delta$) behavior is entirely coincident regardless of $\Delta$.

As expected from the discussion above, the $T$-dependence of the conductance upon increasing the SO coupling $\tilde\Delta$ in the center of the one-electron valley (\fref{fig:zbcondt}(b)), shows clearly a crossover from the $SU(4)$ behavior arising for $\tilde\Delta =0$, to the $SU(2)$ behavior arising asymptotically for $\Delta \gg T_{\mathrm{K}}^{SU(4)}$. This limiting form occurs  in practice for $\tilde\Delta \gtrsim 0.2$
(\ie \ $\Delta / T_\mathrm{K}^{SU(4)} \gtrsim 7$), such that a further increase in $\tilde\Delta$ naturally leaves the $T$-dependence of $G_{\mathrm{c}}^{0}$ unchanged, as seen in the figure.

\subsection{Zero-bias conductance at finite $B$}
\label{sec:zbfiniteb}
We turn now to finite field, fixing the SO coupling $\tilde \Delta$ and considering $G_\mathrm{c}^{0} \equiv G_{\mathrm{c}}^{0}(\tilde{B},N_{\mathrm{g}})$ as a function of the dimensionless gate voltage $N_\mathrm{g} =\tfrac{1}{2}(1-\tfrac{2\epsilon}{U})$ and field $\tilde{B}=\gamma_{\mathrm{s}}B/\Gamma$. We first consider $T=0$, in terms of which the finite-$T$ behavior is readily understood.

\begin{figure}
\includegraphics{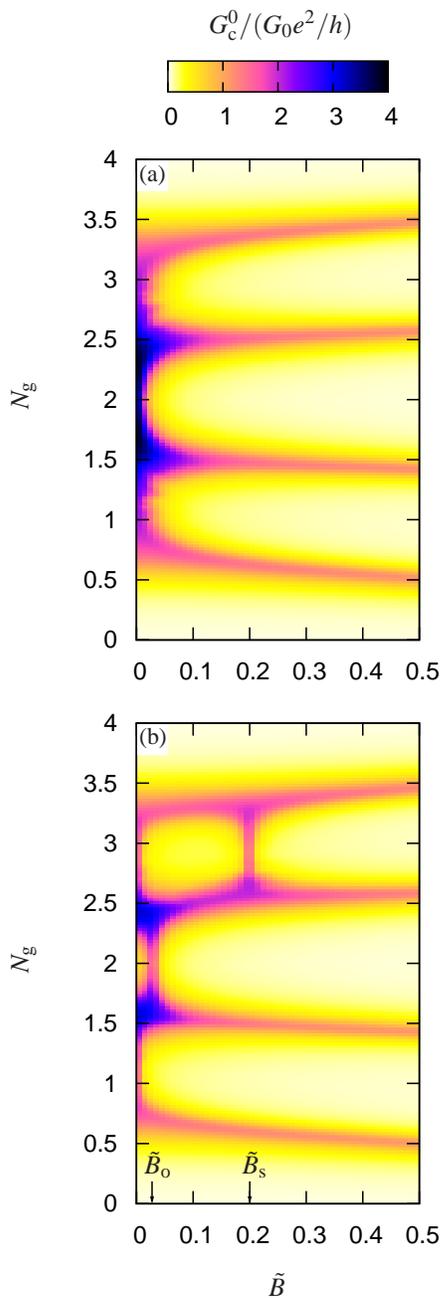}
\caption{\label{fig:condmaps}(Color online.) $T=0$ zero-bias conductance maps, 
$G_\mathrm{c}^{0}/(G_0 e^2/h)$ as a function of field $\tilde B = \gamma_\mathrm{s}B/\Gamma$ and dimensionless gate voltage $N_\mathrm{g}$, for (a) $\tilde\Delta=0$ and  (b) $\tilde\Delta=0.2$. Bare parameters here are $\tilde U =20$
and $\gamma_{\mathrm{o}}/\gamma_{\mathrm{s}} =7$. }
\end{figure}

The $\tilde\Delta = 0$ conductance as a function of field\cite{Choi2005,Rui2006,Busser2007} is shown in \fref{fig:condmaps}(a).
At $B=0$ it has the stepwise form seen in \fref{fig:zbcondv}, with a maximal conductance plateau of $G_\mathrm{c}^{0}/G_0 = 4e^2/h$ in the center of the two-electron valley ($N_\mathrm{g} = 2$) and plateaux of $G_\mathrm{c}^{0}/G_0 \simeq 2e^2/h$ in the centers of the one- and three-electron valleys.
For $B \neq 0$, the associated $SU(4)$ Kondo effects are progressively destroyed: four distinct CB peaks instead emerge, centered along the lines of atomic limit charge degeneracy (\cf\ \fref{fig:cbalnoso}). The $SU(4)$ Kondo behavior at $B=0$ persists over  a finite $B$-range, but is eventually destroyed for 
$\gamma_\mathrm{s} B\gg T_\mathrm{K}^{SU(4)}$.

On introducing a finite SO coupling $\tilde \Delta = 0.2$, the picture changes to that of \fref{fig:condmaps}(b) (\cf\ \fref{fig:cbal}). The Coulomb blockade lines now show the expected kinks at the fields
$B = B_\mathrm{o}$ and $B = B_\mathrm{s}$, while for sufficiently high $B\gg B_\mathrm{s}$, SO coupling is of course negligible and the behavior approaches that of \fref{fig:condmaps}(a). 

\begin{figure}
\includegraphics{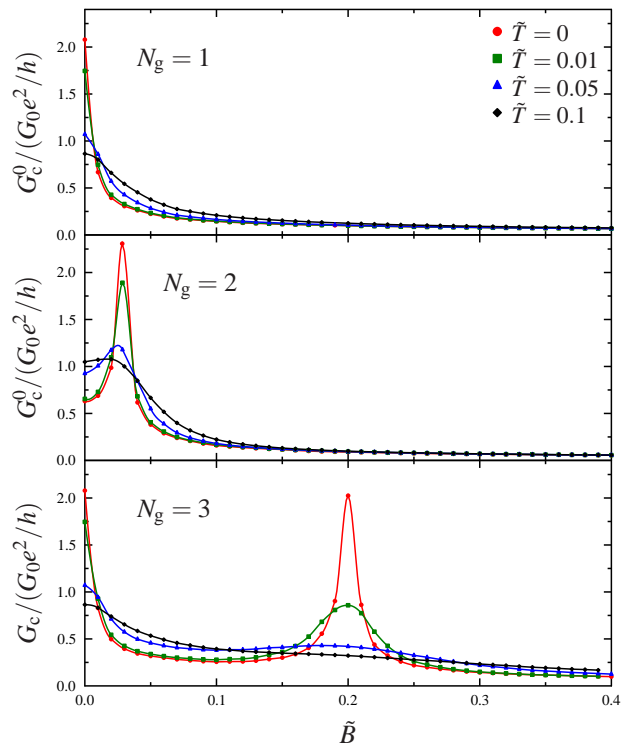}
\caption{\label{fig:cbvalleyt}(Color online.) Slices through the $N_{\mathrm{g}}=1$, $N_{\mathrm{g}}=2$ and $N_{\mathrm{g}}=3$ valleys of \fref{fig:condmaps}(b), showing the evolution of the conductance $G_\mathrm{c}^{0}$ with temperature $\tilde T$ as indicated.} 
\end{figure}
  
  Of primary interest here is the effect of SO coupling on the Kondo physics. As seen earlier, at $B=0$ the SO coupling destroys $SU(4)$ Kondo in the two-electron valley, eliminating the conductance maximum in this region. It also reduces the one- and three-electron zero-field $SU(4)$ Kondo effects to $SU(2)$ 
(the latter apparent in \fref{fig:condmaps}(b) from a clear reduction in the field strength required to destroy the Kondo peak). Although SO coupling thus has a destructive influence on the Kondo effects for $B=0$, it leads as discussed in \sref{sec:atomic} to two finite-field $SU(2)$ Kondo effects (\fref{fig:condmaps}(b)) when level crossings occur in the atomic limit ground states. Both lead to a significant enhancement of the $T=0$ conductance, of order $2e^2/h$ and extending over field ranges $\gamma_\mathrm{s}B\sim T_\mathrm{K}^{SU(2)}$, resulting in marked differences between \fref{fig:condmaps}(a) and \fref{fig:condmaps}(b). 

The effect of temperature on conductance maps is best seen by taking slices through the centers of the CB valleys in the $(N_\mathrm{g},\tilde B)$ plane. The center of the two-electron valley is (by symmetry) $\epsilon = - 3U/2$ for all $B$, while for the one- and three-electron valleys we take the trajectories
\begin{equation}
\label{eq:onemid}
\epsilon = -U/2 + \half\gamma_\mathrm{o}B
\end{equation}
and
\begin{equation}
\label{eq:threemid}
\epsilon = -5U/2 - \half\gamma_\mathrm{o}B
\end{equation}
respectively (which approach the centers of the CB valleys in the large-$B$ limit). 

The results are shown in \fref{fig:cbvalleyt}.
In the two-electron valley the only Kondo effect is the $SU(2)$ Kondo `revived' at the finite field $B=B_\mathrm{o}$ ($\tilde B_{\mathrm{o}} \simeq 0.028$ here).\cite{TieFeng2008} The corresponding $T_{\mathrm{K}}^{SU(2)}\simeq 0.007 \Gamma$, and hence on increasing $T$ the Kondo effect is in essence destroyed by $\tilde T = T/\Gamma = 0.1$. 
Directly analogous comments apply to the zero-field Kondo effects arising in the one- and three-electron valleys;
and, for the latter case, to the additional finite-field $SU(2)$ Kondo effect revived at $B=B_\mathrm{s}$ ($\tilde B_{\mathrm{s}} = 0.2$ here). We also note that the clear SO-induced asymmetry between the one- and three-electron valleys persists even for temperatures $T/\Gamma \sim 0.1$, where the finite-field Kondo peak at $\tilde B_{\mathrm{s}}$ is itself thermally washed out: relatively small though it is, the conductance in the $N_{\mathrm{g}}=3$ valley appreciably exceeds that in the $N_{\mathrm{g}}=1$ valley 
over a wide $\tilde B$-interval.


\subsubsection{Weaker correlations}
\label{sec:wkcoupling}
Thus far we have focussed on the strongly correlated regime where $U/\Gamma \gg 1$. This leads to a pristine separation of energy scales: the CB peaks are separated by many times their widths, and the Kondo scales are exponentially smaller than $\Gamma$. 

\begin{figure}
\includegraphics{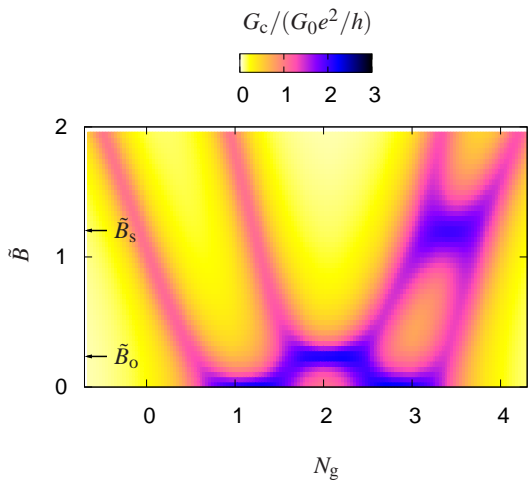}
\caption{\label{fig:condmapwk} (Color online.) As \fref{fig:condmaps}(b), but for a reduced coupling strength $\tilde U=6$, with $\tilde\Delta=1.2$ and $\gamma_\mathrm{o}/\gamma_\mathrm{s}=5$.}
\end{figure}

On moving to a more moderately correlated regime, the energy scales naturally begin to merge, but the essential situation remains the same. An example is shown in \fref{fig:condmapwk}, where now  $\tilde U = 6$ and $\tilde \Delta = 1.2$ (and  $\gamma_\mathrm{o}/\gamma_\mathrm{s}$ has been reduced slightly to $5$).
For $\tilde U = 6$, the resultant~\cite{footnoteTK} zero-field $SU(2)$ Kondo scale for example is 
$T_\mathrm{K}^{SU(2)}\simeq 0.5\Gamma$, and hence $T_\mathrm{K}^{SU(2)}$, 
$\Delta$ and $\Gamma$ all comparable. As seen from \fref{fig:condmapwk}, in comparison to the more strongly correlated \fref{fig:condmaps} this generates a more pronounced asymmetry between the one- and three-electron valleys on increasing $B$, in particular with the CB peaks in the three-electron valley brought closer together.

  The results shown in \fref{fig:condmapwk} agree well with the experimental results of JH~\cite{jarillo2} for the four-electron shell centered on $V_\mathrm{g}\sim 3\mathrm{V}$,~\cite{footnoteBusser} as evident from direct comparison with fig.~2(b) of JH in the interval $2.5\text{V} \lesssim V_\mathrm{g} \lesssim 3.5\text{V}$. And more significantly, they provide a natural explanation for the observations, as arising from the interplay between spin-orbit and Kondo physics.

The bare parameters employed in \fref{fig:condmapwk} are themselves  consistent with the JH experiment. From fig.~2(b) of JH (the $\mathrm{C}_{2}$-$\mathrm{D}_{2}$ line therein) one readily identifies the experimental $B_\mathrm{s} \simeq 3\text{T}$, and likewise the experimental ratio $(\gamma_\mathrm{o}/\gamma_\mathrm{s} =)~B_\mathrm{s}/B_\mathrm{o} \simeq 5$ (from the $\mathrm{C}_{2}$-$\mathrm{D}_{2}$ and $\mathrm{B}_{1}$-$\mathrm{C}_{1}$ lines). Since $B_{\mathrm{s}}=\Delta/\gamma_{\mathrm{s}}$ (\eref{eq:bsdef}), and 
$\gamma_\mathrm{s}/2 = 0.058\text{meV}\text{T}^{-1}$ (we take $g =2$), the experimental $B_{\mathrm{s}}$ gives the SO coupling constant as $\Delta = \gamma_{\mathrm{s}}B_{\mathrm{s}} \simeq 0.35 \text{meV}$ -- which we note is in line with that measured recently in the CNT experiments of [\onlinecite{rmce}] via sequential tunneling spectroscopy at finite bias.

Fitting the atomic limit CB peaks to the JH experimental data gives $U/\Delta\simeq 5$ and hence $U\simeq 2\text{meV}$, in good agreement with the heights of the CB diamonds in fig. 2 of JH; while the ratio $U/\Gamma$ is 
estimated straightforwardly by comparing the widths of the CB peaks to their separation. Finally, the experimental temperature $T = 0.34\text{K} \simeq 0.03\mathrm{meV}$ is sufficiently small compared to the other scales in the problem, that one can set $T=0$ with impunity in the NRG calculations.

As above, we consider the behavior seen in \fref{fig:condmapwk} to be in striking agreement with fig.~2(b) of JH in the region $2.5\text{V} \lesssim V_\mathrm{g} \lesssim 3.5\text{V}$. We also point out that the experiment
deviates from our calculation above a gate voltage $V_{\mathrm{g}}\simeq 3.5\text{V}$. This arises simply because the levels of the \emph{adjacent} four-electron shell in experiment are brought into play by the magnetic field, 
and at sufficiently high $B$ `interfere' with those arising from the shell considered. This is naturally not taken into account in the model (although it would be straightforward to incorporate).

\subsection{Finite bias}
\label{sec:finitebias}

\begin{figure*}
\includegraphics{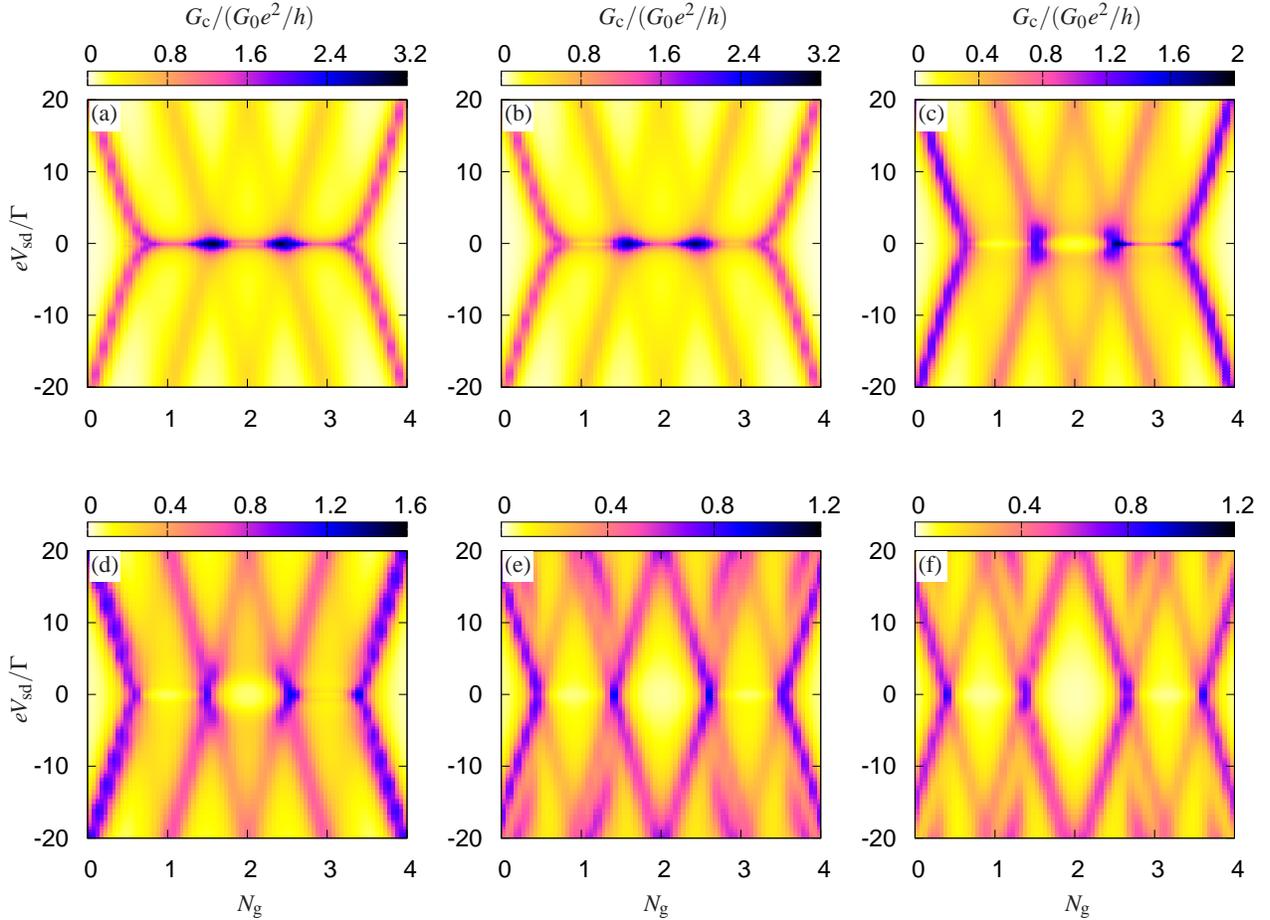}
\caption{\label{fig:cbf} (Color online.) Finite-bias differential conductance, as a function of dimensionless gate voltage $N_{\mathrm{g}}=\tfrac{1}{2}(1-\tfrac{2\epsilon}{U})$ $\propto V_{\mathrm{g}}$ and source-drain bias 
$e V_\mathrm{sd}/\Gamma$; at $T=0$ and for field strengths $\tilde B$ of (a) $0$, (b) $0.04$, (c) $0.2$, (d) $0.4$, (e) $1.0$, (f) $1.4$. The other bare parameters are $\tilde U =20$, $\tilde \Delta=0.2$ and $\gamma_\mathrm{o}/\gamma_\mathrm{s}=5$, such that for (b), $B=B_\mathrm{o}$ and for (c), $B=B_\mathrm{s}$. }
\end{figure*}

So far we have considered the zero-bias conductance as a function of gate voltage, magnetic field and temperature. Experimentally there is another `knob to turn', the source-drain bias voltage $V_{\mathrm{sd}}$. As explained in \sref{sec:theory}, this we handle approximately using \eref{eq:gencond}, which relates the finite-bias conductance to the frequency dependence of the equilibrium single-particle spectra $D_{i\sigma}(\omega)$.  

Our main interest here is how finite fields and SO coupling affect the low-energy Kondo behavior of the conductance. When $\Delta=0=B$, all four $D_{i\sigma}(\omega)$s share a common~\cite{footnotedistinct} $SU(4)$ Kondo resonance in each of the $N_\mathrm{g}=1$, $2$ and $3$ CB valleys. On introducing a finite $B$ and $\Delta$, and thus lowering the symmetry to $U(1)\times U(1)\times U(1)\times U(1)$, each $D_{i\sigma}(\omega)$ instead possesses a distinct Kondo resonance.  On application of a field, these resonances shift away from the Fermi level $\omega =0$ by different amounts, and at high fields in particular the four resonances are sufficiently well separated
that the combined spectrum $\sum_{i,\sigma} D_{i\sigma}(\omega)$ contains four separate peaks.\cite{Choi2005} 
We also emphasise at this point that, despite occasional naive belief to the contrary, the field-induced shifts of the Kondo resonance are \emph{not} simple `Zeeman splittings': they have in fact a non-linear field dependence brought about by the strong interactions on the dot, which can either underestimate or overestimate the Zeeman splitting depending on the strength of the field (see \eg\ [\onlinecite{Logan2001}] and refs therein for a discussion of the $SU(2)$ Anderson model).

Bearing the above in mind, we consider (\fref{fig:cbf}) finite-bias differential conductance maps, $G_\mathrm{c}$ as a function of $V_\mathrm{sd}$ and gate voltage $N_\mathrm{g}$, with each taken at fixed field. Again we start at $T=0$, moving to finite-$T$ later when  comparing to the experiments of M~\cite{FinkelsteinI2007}. Taking the limit $T\to 0$ in \eref{eq:gencond} shows that $G_\mathrm{c}(V_\mathrm{sd})$ is proportional to the average of $\sum_{i,\sigma} D_{i\sigma}(\omega =\pm \tfrac{1}{2}e V_\mathrm{sd})$, \ie\ the (approximate) conductance 
amounts to a symmetrized combination of the total single particle spectrum of the dot.

The $B=0=\Delta$ behavior has been described in [\onlinecite{Anders2008}] (fig.~5 therein). Two distinct features arise: the narrow zero-bias  $SU(4)$ Kondo ridges produced by coherent many-body tunneling, and  finite bias Coulomb blockade diamonds generated by incoherent sequential tunneling. The former occur only below $T$s of order $T_\mathrm{K}^{SU(4)}$ and are likewise destroyed by the source-drain bias when $e V_\mathrm{sd}$ becomes of the same order, while the latter are of width $\sim\max(\Gamma,T)$ and hence rather more robust.

Figure~\ref{fig:cbf} shows the behavior for a finite $\tilde\Delta = 0.2$ at $T=0$, for a range of field strengths.
The $B=0$ conductance is shown in \fref{fig:cbf}(a). The Coulomb blockade diamonds are essentially unchanged from the $\Delta = 0$ limit\cite{Anders2008} since $\Delta\ll\Gamma$, and the form of the zero-bias conductance is as
discussed in relation to \fref{fig:zbcondv}: on switching on $\Delta$, the one- and three-electron valleys 
($N_\mathrm{g}\simeq 1$ and $3$) show $SU(2)$ Kondo effects instead of $SU(4)$, while the conductance in the two-electron valley at $N_\mathrm{g}\simeq 2$ is substantially reduced. We now see from \fref{fig:cbf}(a) that the reduction of the zero-bias conductance in the two-electron valley in fact reflects a splitting of the Kondo resonance in the symmetrized spectrum: two narrow conductance peaks are seen to arise for $N_\mathrm{g}\simeq 2$ when $eV_\mathrm{sd}\sim \pm\Delta$.

On slightly increasing the field to $B=B_\mathrm{o}$, \fref{fig:cbf}(b), the behavior around $eV_\mathrm{sd}\sim 0$ changes. First, the $SU(2)$ Kondo resonances in the one- and three-electron valleys split
(in analogy to the well known behavior of the $SU(2)$ AIM in a magnetic field). 
In the two-electron valley at $B=B_\mathrm{o}$ by contrast, the orbital $SU(2)$ Kondo effect described earlier arises, and as such the splitting of the Kondo resonance here is reduced to zero. 

The next `special' value of the field is $B=B_\mathrm{s}$, shown in \fref{fig:cbf}(c). Here the $SU(2)$ spin Kondo effect takes place in the three-electron valley, whence the splitting of the Kondo resonance seen in \fref{fig:cbf}(b) is reduced to zero. In addition, two  faint `shoulders' at a small finite bias can just be made out. These are the beginnings of the separation of the total spectrum $\sum_{i\sigma}D_{i\sigma}(\omega)$ into four separate components at high field (as mentioned above), which we discuss in more detail below. 

Finally, for $B>B_{\mathrm{s}}$ the atomic limit ground states in all valleys are unchanged with increasing $B$, and the low-energy peak splittings in the $V_{\mathrm{sd}}$-dependence of the conductance all increase monotonically. Moving from (d) to (e) and (f) in \fref{fig:cbf}, we see that once the field becomes of order $\Gamma$ the Kondo peaks simply merge with the CB diamonds (or the Hubbard satellites in single-particle spectra terminology), and the latter themselves begin to split from then on.

\begin{figure}
\includegraphics[scale=1.1]{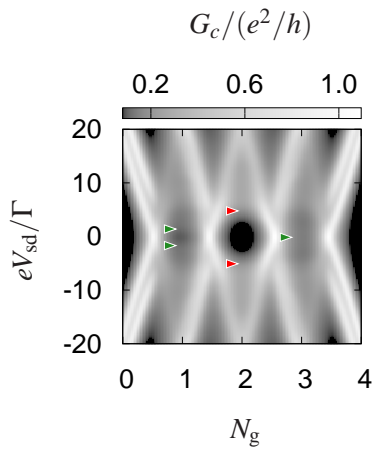}
\caption{\label{fig:cbft} As \fref{fig:cbf} but with $\tilde B=0.7$, $\tilde T = 0.3$, and a colormap chosen to be similar to fig.~2 of [\onlinecite{FinkelsteinI2007}]. The arrows indicate the positions of low-energy peaks, see text. Taking $\Gamma = 0.5 \text{meV}$, the $e V_\mathrm{sd}$ axis extends from $-10$ to $10$ meV, $B\simeq 3\text{T}$ and $T\simeq 2\text{K}$, in agreement with fig.~2 of [\onlinecite{FinkelsteinI2007}].} 
\end{figure}

At this point we make our first comparison with the experiment of M~\cite{FinkelsteinI2007}. Keeping the ratio $U/\Gamma = 20$, we take $\Gamma = 0.5\text{meV}$ such that the resulting $U=10\text{meV}$ is in good agreement with the heights of the CB diamonds in fig.~2 of M. Our choice of $\Delta/\Gamma = 0.2$ then corresponds to $\Delta = 0.1\text{meV}$, which is of the same order of magnitude as that measured 
in another device.\cite{rmce} And we now consider $T/\Gamma = 0.3$ and $\gamma_\mathrm{s} B/\Gamma = 0.7$, 
to be in line with  the experimental temperature and field, $T = 2\text{K}$ and $B = 3\text{T}$. Using a colormap similar to M, and taking $G_0=1$, we obtain \fref{fig:cbft},
which is to be compared with fig.~2 of M.

\begin{figure*}
\begin{center}
\includegraphics{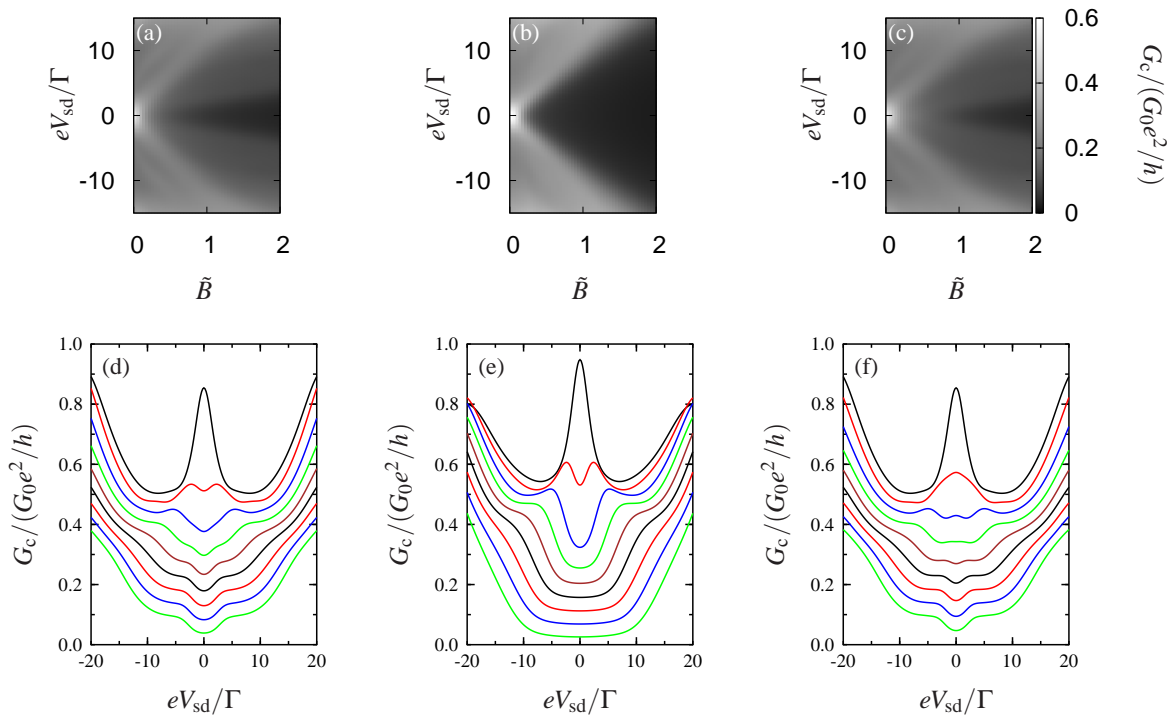}
\end{center}
\caption{\label{fig:valleyvb} Field dependence of the low-energy conductance peaks in the centers of the CB valleys,
for $\tilde T =0.3$. (a), (b) and (c) show $G_\mathrm{c}/(G_0 e^2/h)$ as a function of $\tilde B$ and $eV_{\mathrm{sd}}/\Gamma$, along the centers of the one-, two- and three-electron valleys respectively.
(d), (e) and (f) show slices of the above data taken at fixed fields $\tilde B = 0$, $0.2$, $0.4$, $\ldots$ , $1.6$ (top to bottom). For clarity, the data in (d)--(f) have been shifted vertically by $0.2(1.6 - \tilde B)$ to separate the individual lines.}
\end{figure*}

Our results are in good agreement with the experiment. In particular, we note that the low-energy features identified in the experimental paper\cite{FinkelsteinI2007} are reproduced by the calculations. In the two-electron valley we see a circular region of reduced conductance, while in the one- and three-electron valleys we capture two and one low-energy peak, respectively (marked by arrows in \fref{fig:cbft}). 

On comparing \fref{fig:cbft} to the $T=0$ results in \fref{fig:cbf}, the apparent single peak in the center of the three-electron valley is in fact seen to be a consequence of thermal broadening: At zero temperature in the three-electron valley, a single peak centered at zero-bias occurs only at the special fields of $B=0$ and $B=B_\mathrm{s}$ as explained above (at any other field, this zero-bias peak is always split). While a similar splitting arises also in the one-electron valley, it increases monotonically from $B=0$ and is hence somewhat larger than that of the three-electron valley. As a result, the two peaks in the one-electron valley remain separate at the particular temperature $\tilde T = 0.3$ ($T\simeq 2\mathrm{K}$) used in \fref{fig:cbft}, while the two peaks in the three-electron valley are merged into one. If the experiment had been performed at a sufficiently lower temperature, we would expect two low-energy peaks in both the one- and three-electron valleys.

To compare further with experiment, we now fix the gate voltage to lie at the centers of the one-, two- and three-electron valleys (\cf\ \fref{fig:cbvalleyt}) and consider the conductance as a function of source-drain bias and field. The results are shown in \fref{fig:valleyvb}(a)--(c) as colormaps, and in \fref{fig:valleyvb}(d)--(f) as slices at fixed $\tilde B=0$, $0.2$, $0.4$, $\ldots$, $1.6$. For clarity, the data in \fref{fig:valleyvb}(d)--(f) have been shifted vertically by $0.2(1.6 - \tilde B)$ to separate the individual lines. The figure is to be compared with Fig. 3 of M.

We see in \fref{fig:valleyvb}(a)--(c) the evolution of the low-energy conductance peaks with increasing field, which again agree rather well with experiment. In the one-electron valley (\fref{fig:valleyvb}(a)) the single Kondo peak at $B=0$ is seen to split into the four spectral features highlighted in M;  while, as discussed earlier, in the three-electron valley (\fref{fig:valleyvb}(c)) for sufficiently-small $B$ the two lowest-energy peaks are merged by thermal broadening. For larger $B$ ($\tilde B \gtrsim 1$ here) the two peaks in the three-electron valley do eventually separate in our calculations, as expected on physical grounds. This splitting is difficult to see in the experiment due to the neighboring four-electron shell being brought into play at high field, but should be observable in a similar device with a larger energy separation between shells. 

The destruction of the Kondo effect in the two-electron valley with increasing $B$, as observed in M, is also clearly seen (\fref{fig:valleyvb}(b)), first as a splitting of the $B=0$ Kondo resonance which then rapidly enlarges to leave an almost rectangular-shaped `hole' in the conductance around $V_\mathrm{sd}=0$ (\fref{fig:valleyvb}(e)). 
Note also that the data slices shown in \fref{fig:valleyvb}(d--f) are symmetrical about $V_{\mathrm{sd}}=0$, reflecting the assumption in the calculations of a perfectly symmetrical voltage split between the leads
(\sref{sec:theory}). Their experimental counterparts in fig.\ 3 of M for the one- and two-electron valleys (which are not appreciably affected by `overlap' with a higher shell) are somewhat asymmetrically disposed about $V_{\mathrm{sd}}=0$. This can in fact be reproduced in calculation by parametrising a small degree of asymmetry into the voltage split (without affecting the essential quality of \fref{fig:valleyvb}(a--c)), although we do not pursue it further here.


\section{Conclusion}
\label{sec:conc}
In this paper we have studied the effect of SO coupling in carbon nanotube quantum dots, by applying the NRG to a modified $SU(4)$ Anderson impurity model (AIM). Our main focus has been the case in which the SO coupling is comparable to or exceeds the $SU(4)$ Kondo scale, since here the two effects interplay and compete, leading to a rich range of physical behavior. The differential conductance over a wide parameter space has been calculated as a function of gate voltage, magnetic field and temperature, in order to elucidate the key physics of the model. We have moreover shown that the inclusion of SO coupling accounts for a number of important experimental observations in the works of Jarillo-Herrero \etal\ \cite{jarillo2} and Makarovski \etal\ \cite{FinkelsteinI2007}, the origin of which stems directly from the interplay between SO and Kondo physics.

To conclude, we comment on the suitability of the `pure' $SU(4)$ AIM as a model for carbon nanotube quantum dots.  In [\onlinecite{Anders2008}], experimental data of Makarowski \etal \ at zero-field\cite{FinkelsteinII2007} was found to be in good agreement with the $SU(4)$ Anderson model \emph{without} including SO coupling, over a wide range of $U/\Gamma$. Given the results of the present paper, one naturally asks: why? 

Let us first summarise the experiment. The conductance of the experimental device\cite{FinkelsteinII2007} was measured, at several fixed temperatures, as a function of the applied gate voltage. The latter was swept over a sufficiently wide range that four different electron shells (`Groups I--IV') were brought through the Fermi level, one at a time. A consequence of varying the gate voltage by this relatively large amount 
was that the tunnel couplings to the dot ($\Gamma$) varied from one shell to the next. As a result, the Group I data were described by an $SU(4)$ model with $\Gamma\simeq 0.5\text{meV}$ (and $U/\Gamma=20$), Groups II and III were more consistent with $\Gamma\simeq 1\text{meV}$ ($U/\Gamma=10$), while Group IV had $\Gamma\simeq 2\text{meV}$ ($U/\Gamma=5$).

To understand why these data could be described by the $SU(4)$ model, we note that (a) the experimental `base' temperature was $1.3\text{K} \simeq 0.11 \text{meV}$ and (b) the Kondo scales for Groups II--IV are all in excess of $4\mathrm{K}\simeq 0.36\text{meV}$ (see fig.~3(b) of [\onlinecite{FinkelsteinII2007}]). Assuming the SO coupling 
to be comparable to the value $0.1\text{meV}$ considered above, Groups II--IV \emph{can} then be described by a pure $SU(4)$ Anderson model (for essentially all temperatures), since each has a $T_\mathrm{K}^{SU(4)}$ appreciably
in excess of the SO coupling. While the latter is not the case for the Group I shell~\cite{Anders2008} of [\onlinecite{FinkelsteinII2007}], data were only obtained down to a temperature of order $\Delta$ where, according to \sref{sec:zbb0}, the effects of SO coupling cannot be seen in the zero-field behavior alone. Only by examining the behavior of the Group I shell in a magnetic field (as considered here), or by measuring its zero-field conductance down to a rather lower temperature on the order of $\sim 0.1 \text{K}$ or so, can the effects of SO coupling be observed.

\acknowledgments
{We are grateful to Gleb Finkelstein for stimulating discussions.
MRG, FJ and DEL thank the EPSRC (UK) for financial support, under Grant EP/D050952/1. MRG gratefully acknowledges the Oxford e-Research Centre, OxGrid,~\cite{oxgrid} and the UK National Grid Service for providing computer time. 
FBA acknowledges financial support  from the DFG (Germany), under AN 275/6-1.
}


\begin{thebibliography}{61}
\bibitem[{foo({\natexlab{a}})}]{footnoteEoM}
\bibinfo{note}{It is for example readily shown to break down at low
  energy/temperature, and to violate Fermi liquid behavior.}
\bibitem[{foo({\natexlab{b}})}]{footnoteTK}
\bibinfo{note}{In practice $T_\mathrm{K}^{SU(N)}$ is defined as the temperature
  for which the impurity entropy\cite{kww1} is $\tfrac{1}{2}\ln(N)$, suitably
  below its value for the $SU(N)$ local moment fixed point.}
\end{thebibliography}

\end{document}